\documentclass[pre,twocolumn,showpacs,preprintnumbers,amsmath,amssymb]{revtex4}

\usepackage{graphicx}
\usepackage{dcolumn}
\usepackage{bm}
\usepackage{color}

\begin{document}

\title{Observation of particle pairing in a two-dimensional plasma crystal}

\author{S. K. Zhdanov}
\affiliation{Max Planck Institute for extraterrestrial Physics, D-85741 Garching, Germany}

\author{L. Cou\"edel,}
\affiliation{CNRS, Universit\'e d'Aix-Marseille, PIIM UMR 7345, 13397 Marseille Cedex 20, France}

\author{V. Nosenko}
\affiliation{Max Planck Institute for extraterrestrial Physics, D-85741 Garching, Germany}

\author{H. M. Thomas}
\affiliation{Max Planck Institute for extraterrestrial Physics, D-85741 Garching, Germany}
\email{thoma@mpe.mpg.de}

\author{G. E. Morfill}
\affiliation{Max Planck Institute for extraterrestrial Physics, D-85741 Garching, Germany}

\date{\today}

\begin{abstract}
The observation is presented of naturally occurring pairing of particles and their cooperative drift in a two-dimensional plasma crystal. A single layer of plastic microspheres was suspended in the plasma sheath of a capacitively coupled rf discharge in argon at a low pressure of $1~\text{Pa}$. The particle dynamics were studied by combining the top-view and side-view imaging of the suspension. Cross analysis of the particle trajectories allowed us to identify naturally occurring metastable pairs of particles. The lifetime of pairs was long enough for their reliable identification.
\end{abstract}


\maketitle

\section{Introduction}
\label{intro}

A weakly ionized gas comprising dust or other fine solid particles
is known as a complex, or dusty, plasma
\cite{Ikezi:1986,Morfill:2009,Fortov:2005}. In experimental studies
of complex plasmas the particle size is of a few nanometers to tens
of microns. Immersed into a plasma, the particles charge up and
interact with each other. It is a well-established fact that complex
plasmas are able to self-organize, forming a highly ordered
structure, \emph{plasma crystal}, when the mutual interparticle
interaction energy exceeds significantly their kinetic energy
\cite{Thomas:1966}. In the presence of gravity, a single-layer, or two-dimensional (2D) plasma crystal can form. Last two decades of studies showed that plasma crystals can be exploited as a useful tool to model or at least mimic at a kinetic level many phenomena as diverse as particle and energy transport in solids and liquids, crystal layer plasticity, phase and structural transitions, etc. \cite{Morfill:2009,Fortov:2005}.

In plasma crystals, as in any other crystalline structures, point
defects and dislocations are ubiquitous
\cite{Nosenko:2008,Zhdanov:2011}. They may present an obstacle to
performing some delicate experiments.
Additionally, plasma crystals sometimes suffer from the presence of
extra particles \cite{footnote:1}, which do not belong to the crystalline structure
and can cause local instabilities and disturb the lattice. (Such
particles are sometimes called ``unstable'', ``anomalous'', etc. or
even, addressing their position in the flow of ions with respect to
a particle layer, ``upstream'' or ``downstream'', see, e.g.,
Ref.~\cite{Du:2012}.) On the other hand, they can be successfully used,
as the studies performed recently have shown, as an active agent in
the plasma crystal heating experiments
\cite{Schweigert:1996,Nunomura:2006}, as a convenient practical
diagnostic tool allowing to test in the simplest way the complex
plasma elasticity modules
\cite{Samsonov:1999,Schwabe:2011,Lenaic:2012}, or as a probe of the
plasma electric field distribution \cite{Kretschmer:2005}.

The particles which constitute the main lattice of a crystal
are called \emph{intralayer} particles. This terminology is also used in the granular media \cite{Combs:2008} and colloidal \cite{Patti:2009} physics. The particles located between the layers in multi-layer crystals \cite{Schweigert:1996,Pieper:1996} are naturally called the \emph{interlayer} particles \cite{Patti:2009,Pieper:1996}.

The dynamics of interlayer particles are cardinally different from those of the intralayer particles. For example, the dynamics of a single second-layer Delrin particle \cite{Combs:2008} free to move on top of a granular dimer lattice, or the cooperative permeation of string-like clusters in colloids of rods \cite{Patti:2009} reveal unusual features. In plasma crystal studies, the particles moving in a plane above a single-layer plasma crystal (they were termed \emph{upstream} particles in Ref.~\cite{Du:2012}) reveal elements of ``strange kinetics'' \cite{Shlesinger:1993} such as channeling and leapfrog motion \cite{Du:2012}.

\begin{figure}[htbp]
\centering
\includegraphics[width=0.85\columnwidth]{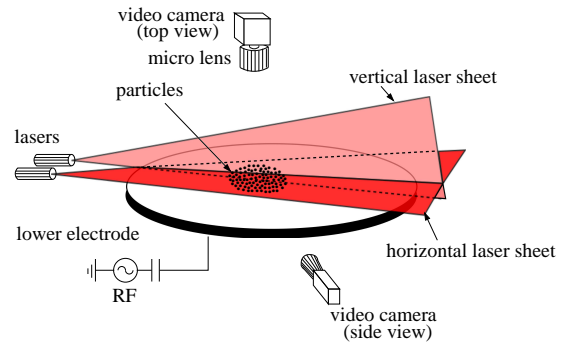}
\caption{(Color online) Sketch of the experimental setup. Plastic microspheres are confined in a stable single layer above the lower rf electrode in a capacitively coupled rf discharge in argon (the top ring-shaped grounded electrode is not shown here). The particle illumination system consists of two orthogonal laser sheets with different wavelengths. The particles are imaged from the top and from the side by two digital cameras equipped with narrow-band interference filters to admit only the respective wavelengths. This setup enables simultaneous recording of the in-plane and out-of-plane particle motion.}
\label{fig:1}
\end{figure}

The interaction of an upstream particle with the plasma crystal located beneath it (downstream of the ion flow) is strongly influenced by the ion wake. An ion wake is a build-up of positive space charge created behind a negatively charged particle by a flow of ions past it \cite{Lampe:2000,Melzer:2001,Steinberg:2001,Ivlev:2003,Lenaic:2009,kroll:2010,Woerner:2012,Morfill:2006,Hyde:2011}. Therefore, the wake-mediated interaction of an upstream particle with the plasma crystal is attraction-dominated \cite{Du:2012}.

A delicate repulsion-attraction balance can result in a strong correlation --
\emph{pairing} of the upstream particle with a neighboring
intralayer one. To some extent, such kind of pairing resembles the
famous Cooper electron pairs when the electron-phonon interactions
produce a strong preference for singlet zero momentum electron pairs
\cite{Cooper:1960}.
Vertical pairing of two identical particles in
the sheath of a rf discharge has been studied in
Ref.~\cite{Steinberg:2001}.

In this paper, we report on the first direct observation of particle
pairing and dragging occurring under natural conditions in a 2D complex plasma. Neither a torque, as in the
``rotating wall'' technique of
Refs.~\cite{Nosenko:2009,Woerner:2012}, nor a laser beam, as in
the laser-dragging experiment of Ref.~\cite{Melzer:2001}, nor any
other method of external manipulation has been used. Using paired
particles as a probe of the mutual interparticle interaction is one
more possibility that is briefly discussed in this paper.

\section{Experimental procedure}
\label{sec.II}

\begin{figure}[htbp]
\centering
\includegraphics[width=0.85\columnwidth]{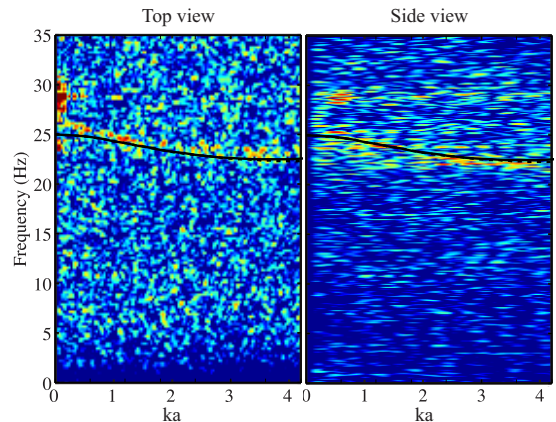}
\caption{(Color) Fluctuation spectra of the out-of-plane particle velocity in a 2D plasma crystal. The spectra were calculated from the top-view experimental movies (using the image intensity-sensitive analysis technique \cite{Samsonov:2005,Lenaic:2009,Lenaic:2012}, left panel) or directly from the side-view movies (right panel). The intensities of the spectra are in arbitrary units (the logarithmic scale spans over two decades). The theoretical dispersion relations \cite{Zhdanov:2009} for the two main crystallographic directions (solid and dashed lines) are very close to each other. A reasonably good agreement of the theoretical and experimental results is evident. The intercept of the out-of-plane phonon spectrum with the frequency axis gives the vertical confinement parameter $f_v$, see Table~\ref{tab:1}. An unrelated excitation at about $29$~Hz is separated from the spectrum by a gap of about $3-5$~Hz.}
\label{fig:2}
\end{figure}

The experiments were performed in a modified version of the Gaseous
Electronics Conference (GEC) rf reference cell \cite{Lenaic:2009}
using argon at a pressure of $1$~Pa and melamine-formaldehyde
microspheres with a diameter of $9.19\pm 0.14~\mu$m, a mass of $m =
6.1 \cdot 10^{-13}~\text{kg}$, and a weight of
$mg=6~\text{pN}$, where $g$ is the free-fall acceleration on
earth. A stylized sketch of the experimental setup is shown in
Fig.~\ref{fig:1}. A weakly ionized plasma is generated by applying a
forward rf power of $15$~W at $13.56$~MHz to the lower disk-shaped rf electrode
(corresponding to the self-bias voltage $V_{\rm dc}=-124$~V). The
microparticles, introduced into the plasma using a dispenser, formed
a stable monolayer confined in the plasma sheath above the rf electrode. Optical ports and windows at the top and the side of
the chamber provide access for the laser illumination and recording
systems. Two digital cameras (a Photron FASTCAM 1024 PCI operating
at $250$ frames per second (fps) and a Basler Ace ACA640-100GM at
$103.56$ fps) recorded the microparticle positions and their
dynamics and provided top- and side-view snapshot sequences
subjected further to a standard particle tracking technique
\cite{Rogers:2007,Williams:2012}. Side-view imaging is usually used in 2D plasma crystal experiments only as a complementary diagnostic. In the present study, we relied on it for our main results.
Therefore, we first verified the side-view data, using the fluctuation spectra of the particle out-of-plane motion, see Fig.~\ref{fig:2}, as a cross-check. Additionally, the side-view camera was used to verify that our experiments were carried out with a (dominantly) single layer of particles.

\section{Complex plasma parameters}
\label{subsec.B}

The plasma crystal parameters were evaluated using a well-developed method based on the particle tracking technique \cite{Williams:2012}. The lattice constant $a$ was obtained from the first peak of the pair correlation function $g(r)$. The neutral gas damping rate was estimated to be $\gamma_E \simeq 1.2~{\rm s}^{-1}$ \cite{Epstein:1924}. The small value of $\gamma_E$ (compared to the characteristic frequency of the plasma crystal) assures weak frictional coupling of the particle dynamics to the ambient gas. Therefore, the particle motion is not overdamped and studying of the naturally occurring waves (fluctuations) can give reliable information about the lattice layer. The values of the particle charge $Q$, interaction range $\kappa=a/\lambda_D$ (where $\lambda_D$ is the screening length), and the vertical confinement parameter $f_v$ were estimated from the fluctuation spectra of particle velocity \cite{Nunomura:2002,Zhdanov:2003,Zhdanov:2009}. These parameters are collected in Table~\ref{tab:1} along with the parameter set adopted for numerical calculations performed for comparison reasons.

\begin{table}[htbp]
\caption {Plasma crystal parameters measured from the top- and side-view recording data as well as the parameter set adopted for numerical calculations.}
\label{tab:1}       
\begin{tabular}{llll}
\hline\noalign{\smallskip}
parameter & top view   & side view  & theory   \\
\noalign{\smallskip}\hline\noalign{\smallskip}
lattice constant, $a$ [$\mu$m] &  $520\pm30^\alpha$ & $530\pm40^\beta$ & 500  \\
interaction range, $\kappa$ & 1.06 &  & 1.06 \\
particle charge, $Q$ [$10^3e$] & $15.0\pm 2.3^\gamma$ &  & 15.3  \\
vertical confinement &  &  &  \\
parameter, $f_v$ [Hz] & $26\pm3$  & $24\pm3$  & 25 \\
longitudinal phonon &  &  &   \\
speed$^\delta$, [mm/s] & $31.0\pm2.2$ & $32\pm4$ & 31  \\
transverse phonon &  &  &   \\
speed$^\delta$, [mm/s] & $6.5\pm1.2$ &  & 6.5  \\
\noalign{\smallskip}\hline
\end{tabular}

$^\alpha$ in the central part of the crystal, measuring $14.6\times14.6\text{mm}^2$; $^\beta$~obtained from a row of particles that was well-aligned with the laser (left half of the top panel in Fig.~\ref{fig:3}); $^\gamma$~for the intralayer particles, assuming no decharging of particles by ion wakes; $^\delta$ in-plane modes.
\end{table}

The experimental fluctuation spectra of the crystal obtained from
either the top-view (TV) or side-view (SV) recording systems are
shown in Fig.~\ref{fig:2}. Although both methods are widely used in complex plasma experimental
studies (see, e.g., \cite{Liu:2010,Lenaic:2012} and the references
therein), a cross-checking diagnostic has never been done before and
the results of the TV and SV observations were never systematically
compared. Below are important points of comparison that are worth to
comment on: (i) the TV- and SV-spectra agree remarkably well
with each other; (ii) the SV-spectra show systematically lower resolution in the wave numbers due to a significantly poorer spatial sampling rate; (iii) the SV-spectrum of the out-of-plane fluctuations is systematically lower (by about $0.5$~Hz) than the TV-spectrum most probably due to the fact that not exactly the same parts of the crystal are analyzed. It is also worth noting that the SV-spectra are not angle-resolved \cite{Liu:2010}. A more detailed comparison will be reported elsewhere.

The electric field in the discharge (pre)sheath is inhomogeneous with the characteristic length given by $L_E=E_0/E_0'=g/(2\pi f_v)^2\simeq0.4~\text{mm}$ under our experimental conditions; here, a balance $Z|e|E_0=mg$ is assumed to be valid.
Note that a dense lattice layer, consisting of highly charged microparticles, itself produces a finite electric field in its vicinity. In our conditions it is not large, though, about one fifth of $E_0$ in the mean-field approximation \cite{Totsuji:2001}.

\section{Direct observation of the interlayer particle collisions}
\label{sec.IV}

\begin{figure}[b]
\centering
\includegraphics[width=0.85\columnwidth]{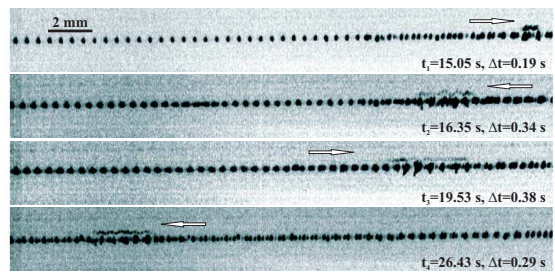}
\caption{
Traces of the fast-moving upstream particles recorded by the side-view camera 
Each panel was obtained by blending a sequence of snapshots, the recording timing is indicated \cite{footnote:2}. The white arrows indicate the direction of motion of upstream particle. The apparent variation of the particle density across the image is due to the domain structure of the crystalline layer and its slow rotation. The illuminating laser light is coming from the left.
Notice a small left grade ($\approx1\%$) of the main layer.
} \label{fig:3}
\end{figure}

Upstream particles spontaneously moving above a 2D plasma crystal were reported for the first time in Ref.~\cite{Du:2012}. Their impact on the dynamics of the crystal layer and some aspects of particle coupling were studied. These fast-moving particles, even if they remain invisible (since they stay outside of the illuminating laser sheet), could be recognized by the appearance of the attraction-dominated Mach cones in the lattice, a signature uniquely manifesting their presence in an experiment. However, a direct observation of the particle pairing process can only be done with the help of a side-view recording system, since the pairs tend to be extended in the vertical direction.

In our experiments, as in Ref.~\cite{Du:2012}, a few upstream particles were wandering quasi-freely on
top of the lattice layer along the channels made by the rows of
ordered intralayer particles. From time to time, encountering a
lattice imperfection blocking the channel, they
strongly scattered and were forced to change the track direction,
then moved again quasi-freely along another newly discovered path,
and so on, covering a large area of the crystal. Usually, this
process took quite a long time.

When an upstream particle happened to move in the vertical laser
sheet, its trace was recorded by the side-view camera, as shown in Fig.~\ref{fig:3}. The travel path of an upstream particle is, on
average, at the height of $\langle\Delta h\rangle\simeq0.2~\text{mm}
\simeq \frac12 L_E$ above the lattice layer (same as estimated in
Ref.~\cite{Du:2012} using a top-view survey). In all cases shown in
Fig.~\ref{fig:3}, the interaction scenario appears to be quite
universal, passing normally through the following well-distinguished
phases: initiation, repulsion, binding, and dragging. When an
upstream particle comes too close to the channel wall or encounters
a point defect, a strong interlayer collision between the top particle and a nearby intralayer
one occurs. The bottom particle drops a little, allowing the
``intruder'' to pass over it. Then the repulsion is apparently
replaced by attraction. The bottom particle starts to behave as if
it was seized by the intruder, tending to be dragged with it. Since
both particles are negatively charged, this is puzzling to some
extent. The newly formed pair continues drifting for a while until
the next strong collision would break it up.

\section{Coupled pairs as quasi-particles}

\begin{figure}[htbp]
\centering
\includegraphics[width=0.85\columnwidth]{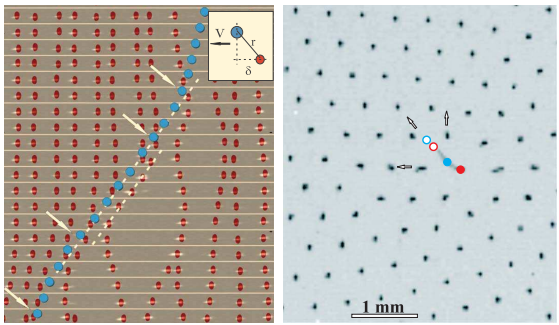}
\caption{(Color) Pairing of an upstream particle with intralayer particles. The left panel shows a space-time diagram assembled from $20$ consecutive side-view images (each approximately $4.3\times 0.8~\text{mm}^2$ in size). Time advances from top to bottom (in the range of $16.38-16.57$~s, see Fig.~\ref{fig:3}), the time step is $0.009646$~s. The cyan and red circles indicate the positions of respectively upstream and intralayer particles. The arrows indicate pairing events. The track of a long-living pair of particles is highlighted by two parallel dashed lines. The inset shows the dragging geometry. The right panel shows the top view of a different dragging event (assembled from $15$ blended images; here, the illuminating laser sheet was shifted upward, which allowed to simultaneously record the upstream and intralayer particles). The filled circles indicate the positions of the upstream (cyan) and intralayer (red) particles in the beginning of a pairing event. The open circles indicate the particle positions $0.06$~s later. The arrows indicate the resultant directions of particle motion.}
\label{fig:4}
\end{figure}

The association of two particles in a pair strongly affects the motion of both particles: They start to accelerate as if the momentum was not conserved during their collision, see Fig.~\ref{fig:4}. This kind of action-counteraction imbalance is not surprising at all keeping in mind the following. First, the binding and subsequent dragging of an intralayer particle, the follower, actually is a direct manifestation of the ion focus (localized positive spatial charge or the ion wake) formed beneath the top particle, which is in the upstream position in the pair. A negatively charged bottom particle \cite{footnote:4} is attracted by the  ion focus while it is repelled by the negatively charged top particle \cite{Morfill:2009,Du:2012}.
At the same time, the bottom particle continues to repel the top one whence accelerating it \cite{Schweigert:1996}. The forces working to produce this motion are the plasma forces \cite{Zhdanov:2005a}.

Newly formed pairs behave as \emph{quasi-particles}, which are (roughly) double-charged compared to the individual particles in the monolayer. This helps them to permeate through the lattice and to find an optimal path, e.g., inside a channel formed by the lattice particles, as the example shown in Fig.~\ref{fig:4} (right panel) demonstrates.

Upstream particles move non-uniformly along their trajectories. For instance, in Fig.~\ref{fig:4}
(left panel) the velocity of such a particle is
about $7~\text{mm/s}$ in the beginning and in the end of the travel
path. However, it is more than twice larger, $\langle
V\rangle\simeq18~\text{mm/s}$, when the particle becomes coupled,
forming a close pair. This acceleration is due to the horizontal
projection of the repulsion force between the coupled particles that
is not completely canceled out. The average distance between the
particles in the pair is $r\simeq0.36~\text{mm}$, its horizontal
projection (dragging distance) is $\delta\simeq0.19~\text{mm}$.
On average, the dragged particle in the pair is kept at the height
$\langle\Delta h\rangle_{\rm drag} \approx 40~\mu\text{m}$ below the
monolayer equilibrium position, experiencing therefore an extra
force of external confinement. This gives a useful estimate of the
$z$-component of the inter-pair repulsion force pressing it down:
$\langle F_z\rangle/mg=\langle\Delta h\rangle_{\rm drag}/L_E \approx
10\%$.

Given the approximately constant velocity of the pair, it is straightforward to roughly estimate the $x$-component of the dragging force: $\langle F_x\rangle/mg \approx 2\gamma_E<V>/g \simeq 0.4\%$. It is about $25$ times weaker compared to the vertical $z$-component, in good agreement with that measured in Ref.~\cite{Du:2012}. Following Refs.~\cite{Woerner:2012,Fink:2012}, the coupling between the particles in a pair can be conveniently interpreted through Hooke's spring constant. Introduced by the relationship $\langle F_x\rangle=k\delta$, where $\delta$ is the dragging distance, it is $k\approx 900~\text{eV/mm}^2$, noticeably well in line with that reported in Refs.~\cite{Woerner:2012,Fink:2012}.

It is also worth noting that the lifetime of a pair is short, e.g., about $0.06~\text{s}$ for the pairs shown in Fig.~\ref{fig:4}. Their formation time (as well as the decomposition time) is even shorter, about $0.01-0.02~\text{s}$. Therefore, these processes are controlled by much stronger coupling forces producing accelerations of the order of $50-100~\text{cm/s}^2$, according to our estimates.

\section{Conclusion}

We have observed for the first time the spontaneously forming mobile pairs of coupled particles in a 2D plasma crystal. This phenomenon is different from previously reported channeling \cite{Du:2012} or ``classical tunneling'' \cite{Morfill:2006}. This observation was made possible by combined top- and side-view imaging of the dust particle suspension. We argue that the apparent self-acceleration of a particle pair is a direct consequence of the plasma wake effect. These naturally-occurring mobile pairs are metastable. They are, however, long-living enough for their reliable detection under our experimental conditions. The pairs we reported on in the present paper were formed by particles located initially at different heights. This helped to initialize the pairing process, because  the mutual wake-mediated interaction was easily activated in this case. It is not strictly necessary for the particles to be \emph{initially} at different heights, though. The pairing of particles is also possible, for instance, in the experimental situations when their vertical displacement becomes relatively large, thus enhancing the mutual wake-mediated interaction. Particle pairing is of primary significance in experimental studies of the later stages of the wake-mediated melting \cite{Lenaic:2009}, as our preliminary observations have demonstrated.

\begin{acknowledgments}

This work was supported by the European Research Council under the European Union's Seventh Framework Programme (FP7/2007-2013) / ERC Grant agreement 267499 and by the French-German PHC PROCOPE program (Project 28444XH).

\end{acknowledgments}


\begin{thebibliography}{99}
\bibitem{footnote:2} The standard blend function of Mathcad 14.0 was used. Blending is useful in substracting and rebiasing backgrounds.
\end{thebibliography}
\end{document}